# Role of the superposition principle for enhancing the efficiency of the quantum-mechanical Carnot engine

Sumiyoshi Abe and Shinji Okuyama

*Department of Physical Engineering, Mie University, Mie 514-8507, Japan*

**Abstract**  A role of the superposition principle is discussed for the quantum-mechanical Carnot engine introduced by Bender, Brody, and Meister [J. Phys. A **33**, 4427 (2000)]. It is shown that the efficiency of the engine can be enhanced by superposition of quantum states. A finite-time process is also discussed, and the condition of the maximum power output is presented. Interestingly, the efficiency at the maximum power is lower than that without superposition.

The superposition principle is at the heart of quantum mechanics. It plays crucial roles in various applications of quantum mechanics. For example, as well known in quantum information, it enables a quantum computer to operate much faster than a classical one [1]. Quantum entanglement, which is also a key concept in quantum information and highlights how the quantum world is different from the classical world, has its origin in this principle. Thus, although the principle itself is simply associated with the linear structure, its significance is immense.

In this paper, we discuss a novel role of the superposition principle. What is considered here is concerned with the quantum-mechanical Carnot engine. We show by employing a simple engine model that superposition of states can significantly enhance the efficiency of the engine. Furthermore, we also discuss finite-time thermodynamics to derive the condition for achieving the maximum power output of the engine. Quite interestingly, the value of the efficiency at the maximum power is smaller than that without superposition.

The quantum-mechanical Carnot engine we are going to study is of the type presented by Bender, Brody, and Meister [2]. It is an analog of the classical Carnot engine, but no heat baths are involved. It is actually a simple two-state model of a single particle confined in a one-dimensional infinite potential well, the width of which can move. The authors of Ref. [2] have devised a reversible process by controlling the quantum states and the potential width. In particular, they have identified the pure-state quantum-mechanical analogs of isothermal and adiabatic processes in thermodynamics, and have shown that remarkably, it is possible to extract the work from such a system. Then, they have found the efficiency of the engine to be given by

$$\eta = 1 - \frac{E_L}{E_H}, \tag{1}$$

where $E_H$ ($E_L$) is the expectation value of the system Hamiltonian along an analog of the isothermal process at high (low) "temperature". This intriguing similarity between pure-state quantum mechanics and thermodynamics has recently been further elaborated in Ref. [3] in view of the micro-macro correspondence. What is essential there is to observe the similarity of the expectation value of the quantum-mechanical Hamiltonian, $H$, in a certain state, $|\psi\rangle$, i.e., $E = \langle \psi | H | \psi \rangle$, to the internal energy. Its change along a "process" is given by $\delta E = (\delta \langle \psi |) H | \psi \rangle + \langle \psi | \delta H | \psi \rangle + \langle \psi | H (\delta | \psi \rangle)$. Therefore, if $(\delta \langle \psi |) H | \psi \rangle + \langle \psi | H (\delta | \psi \rangle)$ and $-\langle \psi | \delta H | \psi \rangle$ are respectively identified with the analogs of the changes of the quantity of heat, $\delta' Q$, and work, $\delta' W$, then the analog of the first law of thermodynamics is established:

$$\delta' Q = \delta E + \delta' W. \tag{2}$$

Let $\{|u_n\rangle\}_n$ be the set of the eigenstates of $H$ satisfying the stationary Schrödinger equation, $H | u_n \rangle = E_n | u_n \rangle$ and be assumed to form a complete orthonormal system. $|\psi\rangle$ is expressed as a superposition of these states: $|\psi\rangle = \sum_n c_n | u_n \rangle$, where the expansion coefficients satisfy the normalization condition, $\sum_n |c_n|^2 = 1$. Then, we have the following analogies:

$$\sum_n E_n \, \delta |c_n|^2 \leftrightarrow \delta' Q, \tag{3}$$

$$\sum_n \delta E_n |c_n|^2 \leftrightarrow -\delta'W . \tag{4}$$

Here, it is important to note that to realize the thermodynamiclike situation, the time scale of the change of the quantum state should be much larger than the characteristic dynamical scale, $\sim \hbar/E$. This requirement allows one to apply the adiabatic scheme in the above-mentioned similrity. That is, the change of $|\psi\rangle$ is dominantly described by the change of the expansion coefficients [2].

Before developing our discussion, it seems appropriate to succinctly summarize the basic points of the quantum-mechanical Carnot engine proposed in Ref. [2]. The Hamiltonian $H$ is of a particle with mass $m$ confined in the one-dimensional infinite potential well with width $L$. The stationary Schrödinger equation, $H|u_n\rangle = E_n|u_n\rangle$, under the obvious boundary conditions yields the energy eigenvalues, $E_n = n^2\pi^2\hbar^2/(2mL^2)$ ($n = 1, 2, 3, ...$). The two-state model in Ref. [2] employs the ground and first excited states, $|u_1\rangle$ and $|u_2\rangle$. The reversible cycle, $A \to B \to C \to D \to A$ is constructed as in Fig. 1. (i) During the process $A \to B$, the state changes from $|u_1\rangle$ to $|u_2\rangle$. In between, it is a superposed state, $a_1(L)|u_1\rangle + a_2(L)|u_2\rangle$, provided that the normalization condition, $|a_1(L)|^2 + |a_2(L)|^2 = 1$, is satisfied. And $E \equiv E_H = [\pi^2\hbar^2/(2mL^2)][4 - 3|a_1(L)|^2]$ is kept unchanged during the expansion, $L_A \to L_B$, in analogy with an isothermal process in classical thermodynamics. From the conditions, $a_1(L_A) = 1$ and $a_1(L_B) = 0$, it follows that $L_B = 2L_A$ and $|a_1(L)|^2 = (4 - L^2/L_A^2)/3$. The force, which is the one-dimensional pressure, is calculated to be $f_{AB}(L) = -\sum_{n=1,2}(\partial E_n/\partial L)|a_n(L)|^2$ $= \pi^2\hbar^2/(mL_A^2 L)$, which decreases as $L^{-1}$. Then, the work is given by $W_{AB} = \int_{L_A}^{L_B=2L_A} dL\, f_{AB}(L) = [\pi^2\hbar^2/(mL_A^2)]\ln 2$, which is the analog of the amount of heat absorbed by the system from the high-temperature heat bath. Therefore, we write, $W_{AB} = Q_H$. (ii) During $B \to C$, $\delta'Q = 0$ in analogy with an adiabatic process in classical thermodynamics, implying that the system stays in $|u_2\rangle$. The force is given by $f_{BC}(L) = -\partial\langle u_2|H|u_2\rangle/\partial L = 4\pi^2\hbar^2/(mL^3)$, which decreases as $L^{-3}$. The work is calculated to be $W_{BC} = \int_{L_B=2L_A}^{L_C} dL\, f_{BC}(L) = (2\pi^2\hbar^2/m)[1/(4L_A^2) - 1/L_C^2]$. (iii) $C \to D$ is in analogy with a process of the isothermal compression. The state changes from $|u_2\rangle$ to $|u_1\rangle$. In between, it is $b_1(L)|u_1\rangle + b_2(L)|u_2\rangle$ with $|b_1(L)|^2 + |b_2(L)|^2 = 1$, and $E \equiv E_L = [\pi^2\hbar^2/(2mL^2)]\left[4 - 3|b_1(L)|^2\right]$ is kept unchanged as in (i). From the conditions $b_1(L_C) = 0$ and $b_1(L_D) = 1$, it follows that $L_D = L_C/2$ and $|b_1(L)|^2 = 4(1 - L^2/L_C^2)/3$. The force and work are given by $f_{CD}(L) = 4\pi^2\hbar^2/(mL_C^2 L)$ and $W_{CD} = -[4\pi^2\hbar^2/(mL_C^2)]\ln 2$, respectively. In particular, $W_{CD}$ is the analog of the amount of heat absorbed by the low-temperature heat bath from the system. Therefore, we write $W_{CD} = -Q_L$. (iv) Finally, during $D \to A$, $\delta'Q = 0$ as in (ii), and the system stays in $|u_1\rangle$. Using $L_D = L_C/2$, the force and work are found to be given by $f_{DA}(L) = -\partial\langle u_1|H|u_1\rangle/\partial L = \pi^2\hbar^2/(mL^3)$ and $W_{DA} = (2\pi^2\hbar^2/m)[1/L_C^2 - 1/(4L_A^2)]$, respectively. It should be noted that $W_{DA} = -W_{BC}$.

Therefore, during the cycle $A \rightarrow B \rightarrow C \rightarrow D \rightarrow A$, the total amount of the work done is $W = W_{AB} + W_{BC} + W_{CD} + W_{DA} = (\pi^2 \hbar^2 / m)(1/L_A^2 - 4/L_C^2) \ln 2$. Thus, the efficiency of the engine, $\eta = W/Q_H = 1 - Q_L/Q_H$ [which is identical to Eq. (1)] is calculated to be

$$\eta = 1 - 4 \left( \frac{L_A}{L_C} \right)^2. \tag{5}$$

In order for the efficiency to be nonnegative, the condition

$$\frac{L_C}{L_A} \geq 2 \tag{6}$$

has to be satisfied.

The expression for the efficiency in Eq. (5) is actually more appropriate than that in Eq. (1), since the latter is written in terms of the expectation values of the Hamiltonian that explicitly depend on the quantum states. (Recall that the Carnot efficiency in classical thermodynamics does not depend on the states of a working material.) For further considerations about this point, see the discussion following Eq. (25) below.

Now, we address ourselves to the problem if the efficiency in Eq. (5) can be improved by superposing the states. For this purpose, instead of the eigenstates, $|u_1\rangle$ and $|u_2\rangle$, we consider their superposed states:

$$|\psi_1\rangle = c_1 |u_1\rangle + c_2 |u_2\rangle, \tag{7}$$

$$|\psi_2\rangle = d_1 |u_1\rangle + d_2 |u_2\rangle, \tag{8}$$

provided that the coefficients satisfy the normalization conditions, $|c_1|^2 + |c_2|^2 = 1$ and $|d_1|^2 + |d_2|^2 = 1$. Note that *the coefficients do not depend on the potential width*, L. Let us consider the cycle in Fig. 1, but now the energy eigenstates, $|u_1\rangle$ and $|u_2\rangle$, are replaced by the superposed states, $|\psi_1\rangle$ and $|\psi_2\rangle$, respectively. (I) Initially at A, the system is prepared in the state, $|\psi_1\rangle$, and changes its state to $|\psi_2\rangle$ at B. During $A \to B$, the state is generically given by $a_1(L)|\psi_1\rangle + a_2(L)|\psi_2\rangle$. [Here, we are using the same notation for the coefficients as in (i) above, but they are not identical, in general.] The expectation value of the Hamiltonian with respect to this state is

$$E = \left[ |a_1(L)|^2 \left( |c_1|^2 + 4|c_2|^2 \right) + |a_2(L)|^2 \left( |d_1|^2 + 4|d_2|^2 \right) + a_1^*(L) a_2(L) \left( c_1^* d_1 + 4 c_2^* d_2 \right) \right.$$

$$\left. + a_1(L) a_2^*(L) \left( c_1 d_1^* + 4 c_2 d_2^* \right) \right] \pi^2 \hbar^2 / (2 m L^2),$$ which is kept constant during the process. Therefore,

$$|a_1(L)|^2 \left( |c_1|^2 + 4|c_2|^2 \right) + |a_2(L)|^2 \left( |d_1|^2 + 4|d_2|^2 \right)$$
$$+ a_1^*(L) a_2(L) \left( c_1^* d_1 + 4 c_2^* d_2 \right) + a_1(L) a_2^*(L) \left( c_1 d_1^* + 4 c_2 d_2^* \right) = \alpha L^2, \tag{9}$$

where $\alpha$ is a positive constant. The conditions, $a_1(L_A) = 1$ and $a_1(L_B) = 0$, fix the value of $\alpha$ to be

$$\alpha = \frac{4 - 3|c_1|^2}{L_A^2} = \frac{4 - 3|d_1|^2}{L_B^2} = \frac{\sqrt{\left(4 - 3|c_1|^2\right)\left(4 - 3|d_1|^2\right)}}{L_A L_B}. \tag{10}$$

Also, since it is an expansion process, the condition

$$\frac{L_B}{L_A} = \sqrt{\frac{4-3|d_1|^2}{4-3|c_1|^2}} > 1, \tag{11}$$

that is, $|c_1| > |d_1|$, should be satisfied. The force and work done are obtained as follows:

$$f_{AB}(L) = \left(4-3|c_1|^2\right)\frac{\pi^2\hbar^2}{mL_A^2 L}, \tag{12}$$

$$W_{AB} \equiv Q_H = \left(4-3|c_1|^2\right)\frac{\pi^2\hbar^2}{mL_A^2}\ln\frac{L_B}{L_A}, \tag{13}$$

respectively. (II) During $B \to C$, the system remains in $|\psi_2\rangle$. The force and work are obtained as

$$f_{BC}(L) = -\frac{\partial}{\partial L}\langle\psi_2|H|\psi_2\rangle, \tag{14}$$

$$W_{BC} = \frac{\pi^2\hbar^2}{2m}\left(\frac{4-3|c_1|^2}{L_A^2} - \frac{4-3|d_1|^2}{L_C^2}\right), \tag{15}$$

respectively, where Eq. (11) has been used in Eq. (15). (III) During $C \to D$, the system changes its state from $|\psi_2\rangle$ to $|\psi_1\rangle$. In-between, it is in $b_1(L)|\psi_1\rangle + b_2(L)|\psi_2\rangle$. The expectation value of the Hamiltonian with respect to this state is calculated to be

$$E = \left[|b_1(L)|^2\left(|c_1|^2 + 4|c_2|^2\right) + |b_2(L)|^2\left(|d_1|^2 + 4|d_2|^2\right) + b_1^*(L)b_2(L)\left(c_1^* d_1 + 4c_2^* d_2\right)\right.$$

$$+b_1(L)b_2^*(L)\left(c_1 d_1^* + 4c_2 d_2^*\right)\right]\pi^2\hbar^2/(2mL^2),$$ which is kept constant during the process as in (I). Therefore,

$$\begin{aligned}&|b_1(L)|^2\left(|c_1|^2+4|c_2|^2\right)+|b_2(L)|^2\left(|d_1|^2+4|d_2|^2\right)\\&+b_1^*(L)b_2(L)\left(c_1^* d_1+4c_2^* d_2\right)+b_1(L)b_2^*(L)\left(c_1 d_1^*+4c_2 d_2^*\right)=\beta L^2,\end{aligned} \quad (16)$$

where $\beta$ is a positive constant. The conditions, $b_1(L_C)=0$ and $b_1(L_D)=1$, fix $\beta$ to be

$$\beta = \frac{4-3|d_1|^2}{L_C^2} = \frac{4-3|c_1|^2}{L_D^2} = \frac{\sqrt{\left(4-3|c_1|^2\right)\left(4-3|d_1|^2\right)}}{L_C L_D}. \quad (17)$$

Also, since it is a compression process, the following condition should hold

$$\frac{L_D}{L_C} = \sqrt{\frac{4-3|c_1|^2}{4-3|d_1|^2}} < 1, \quad (18)$$

which is, in fact, consistent with Eq. (11). The force and work are found to be

$$f_{CD}(L) = \left(4-3|d_1|^2\right)\frac{\pi^2\hbar^2}{mL_C^2 L}, \quad (19)$$

$$W_{CD} \equiv -Q_L = -\left(4-3|d_1|^2\right)\frac{\pi^2\hbar^2}{mL_C^2}\ln\frac{L_C}{L_D}, \quad (20)$$

respectively. Note that Eqs. (11) and (18) imply

$$\frac{L_B}{L_A} = \frac{L_C}{L_D} > 1. \tag{21}$$

(IV) Finally, during $D \to A$, the system remains in $|\psi_1\rangle$ to complete a cycle. The force and work are given by

$$f_{DA}(L) = -\frac{\partial}{\partial L}\langle\psi_1|H|\psi_1\rangle, \tag{22}$$

$$W_{DA} = \frac{\pi^2\hbar^2}{2m}\left(\frac{4-3|d_1|^2}{L_C^2} - \frac{4-3|c_1|^2}{L_A^2}\right) = -W_{BC}, \tag{23}$$

respectively, provided that Eq. (18) has been used in Eq. (23). Thus, the total amount of the work done is found to be

$$W = Q_H - \left(4-3|d_1|^2\right)\frac{\pi^2\hbar^2}{mL_C^2}\ln\frac{L_C}{L_D}, \tag{24}$$

where $Q_H$ is given in Eq. (13).

With the help of Eq. (21), the result is obtained for the efficiency as follows:

$$\eta = 1 - \frac{4-3|d_1|^2}{4-3|c_1|^2}\left(\frac{L_A}{L_C}\right)^2. \tag{25}$$

There is an important point, here. If Eqs. (11) and (18) are used, then this efficiency is rewritten as $\eta = 1 - (L_B/L_C)^2$ or $\eta = 1 - (L_A/L_D)^2$, in which effects of superposition of the states apparently disappear. However, these two expressions are unphysical. In the Carnot cycle, it is essential to start from the state $A$, as pointed out by Clapeyron [4,5]. Therefore, $\eta$ should be expressed in terms of the initial value, $L_A$. Then, the question is, among $\{L_B, L_C, L_D\}$, with which $L_A$ should be combined in $\eta$. The answer is $L_C$, as in Eq. (25). The reason is as follows. The pairs $(L_A, L_B)$ and $(L_D, L_A)$ contain information merely on the single "isothermal" expansion and "adiabatic" compression, respectively. $(L_A, L_C)$ is the one and only physical pair that contains information on the "isothermal" processes at both "high and low temperatures". There exists another reason why the efficiency should be expressed in terms of the pair $(L_A, L_C)$. Consider the total amount of movement of the potential wall (i.e., width) during one cycle, $L_{\text{total}}$. This quantity has to be completely specified by the geometric configuration of the cycle and should be independent of the properties of the quantum states. Clearly, it reads

$$L_{\text{total}} = (L_B - L_A) + (L_C - L_B) + (L_C - L_D) + (L_D - L_A)$$

$$= 2(L_C - L_A), \qquad (26)$$

which is twice the capacity of the engine. That is, $(L_A, L_C)$ is the pair, which geometrically characterizes the cycle independently of the quantum states.

The result in Eq. (25) highlights how the efficiency can be enhanced by superposition of the states. Clearly, the value in Eq. (5) is recovered in the special case when $c_1 = 1$ and $d_1 = 0$. The condition in Eq. (11) leads to

$$|c_1| > |d_1|, \qquad (27)$$

from which the range of the efficiency is found to be

$$1 - 4\left(\frac{L_A}{L_C}\right)^2 \leq \eta < 1 - \left(\frac{L_A}{L_C}\right)^2. \tag{28}$$

Therefore, we see that *the efficiency in Eq. (5) is minimum*. It should be noted however that the supremum of $\eta$ is of no physical interest, since it implies that process (I) is an infinitesimal expansion and process (III) is an infinitesimal compression, as can be seen from Eqs. (11) and (18).

The above observation indicates that maximization of the efficiency does not yield any physically meaningful consequences. This is in marked contrast to the context of the classical Carnot cycle. So, instead of maximizing the efficiency, here we consider maximization of the power output in analogy with finite-time thermodynamic. This discussion casts light on the recent work in Ref. [6].

Let $v(t)$ and $\tau$ be the speed of the change of the potential width, $L$, and the cycle time, respectively. $v(t)$ should be so small that the adiabatic scheme mentioned earlier can be valid. Then, the total amount of movement, $L_{\text{total}}$, in Eq. (26) is

$$L_{\text{total}} = 2(L_C - L_A) = \int_0^\tau dt\, v(t) = \bar{v}\,\tau, \tag{29}$$

where $\bar{v}$ is the average speed. Therefore, the cycle time is given by

$$\tau = \frac{2}{\bar{v}}(L_C - L_A). \tag{30}$$

Rewriting Eq. (24) as

$$W = \left[\left(4-3|c_1|^2\right)\frac{\pi^2\hbar^2}{2mL_A^2} - \left(4-3|d_1|^2\right)\frac{\pi^2\hbar^2}{2mL_C^2}\right]\ln\frac{4-3|d_1|^2}{4-3|c_1|^2}, \tag{31}$$

we have the following expression of the power output:

$$P = \frac{W}{\tau}$$

$$= \frac{\bar{v}}{2L_A^3}\left(4-3|c_1|^2\right)\frac{\pi^2\hbar^2}{2m}f(r,\rho), \tag{32}$$

where

$$f(r,\rho) = \frac{r^2-\rho}{r^3-r^2}\ln\rho, \tag{33}$$

with

$$r = \frac{L_C}{L_A}, \tag{34}$$

$$\rho = \frac{4-3|d_1|^2}{4-3|c_1|^2}. \tag{35}$$

Positivity of $f(r,\rho)$ requires

$$r > \sqrt{\rho} \qquad (1 < \rho \leq 4). \tag{36}$$

Our task is to maximize P with given "initial" values, $L_A$ and $c_1$, at A. Accordingly, we maximize $f(r,\rho)$. In Fig. 2, we present a plot of this function. It has the global maximum. The conditions, $\partial f(r,\rho)/\partial r = 0$ and $\partial f(r,\rho)/\partial \rho = 0$, lead to $r^3 - 3\rho r + 2\rho = 0$ and $\rho \ln \rho - r^2 + \rho = 0$, respectively. The one and only solution of these coupled equations is

$$(\tilde{r}, \tilde{\rho}) = (2.95..., 3.75...), \tag{37}$$

which gives rise to the following value of the efficiency:

$$\tilde{\eta} = 1 - \tilde{\rho}\left(\frac{1}{\tilde{r}}\right)^2$$

$$= 0.56.... \tag{38}$$

Interestingly, this value is slightly smaller than that analytically obtained for the original quantum-mechanical Carnot engine without superposition of the states [6]: $1 - 1/[4\cos^2(2\pi/9)] = 0.573\,977\,952...$.

In conclusion, we have studied a role of the superposition principle for the two-state quantum-mechanical Carnot engine. We have shown that the efficiency of the engine can be enhanced by superposition of the states. We have also discussed the condition of the maximum power output and have found that the corresponding value of the efficiency is slightly lower than that without superposition of the states.

In the present work, the infinite potential well was employed for constructing the engine as in Ref. [2]. Accordingly, the efficiency is expressed in the forms in Eqs. (1) and (5) [or a generalization of the latter in Eq. (25)]. Since an analog of the second law of thermodynamics does not exist in pure-state quantum mechanics [6], these expressions of the efficiency do not have universal meanings. In fact, it can be shown that the form of the efficiency depends on the shape of a potential. Therefore, it is necessary to seek universal features of other kinds. In this sense, the condition of the maximum power output will give a key point, as indicated in Ref. [6]. Further studies in such a direction will contribute to a deeper understanding of similarity between quantum mechanics and thermodynamics considered in Ref. [3].

**ACKNOWLEDGMENT**

S. A. was supported in part by a Grant-in-Aid for Scientific Research from the Japan Society for the Promotion of Science.

______________________________

# Figure Captions

FIG. 1   The quantum-mechanical Carnot cycle depicted in the plane of the width, $L$, and force, $f$.

FIG. 2   Plot of $f(r,\rho)$ with respect to $(r,\rho)$. All quantities are dimensionless.

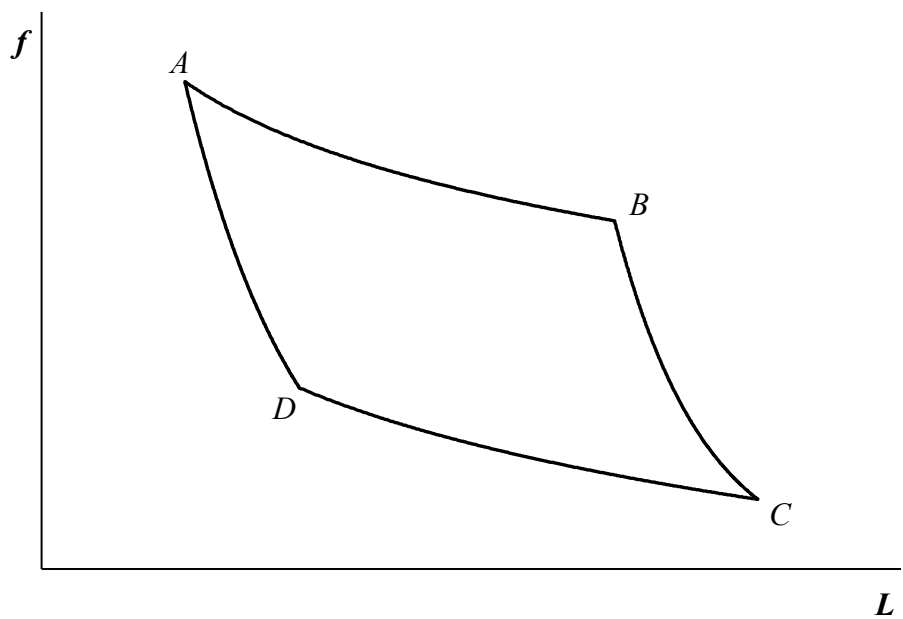

Figure 1

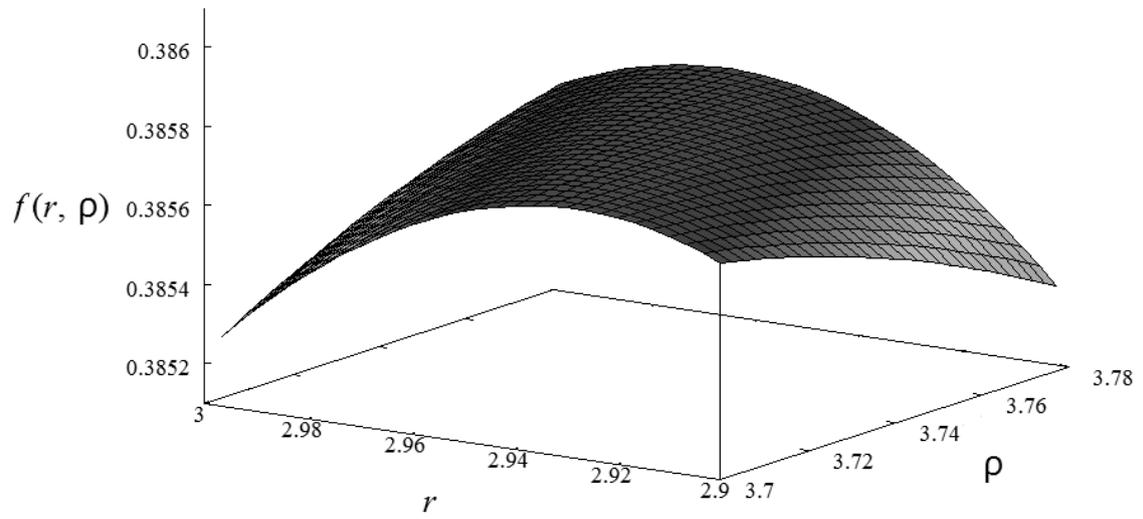

Figure 2